\documentclass{luxenote}
\usepackage[symbol]{footmisc}
\usepackage{tipa}
\usepackage{csquotes}

\title{Test of a partly instrumented highly compact and granular electromagnetic calorimeter in an electron beam of 1 to 6 GeV }

\author{Adri\'an Irles on behalf of the FCAL Collaboration\footnote[1]{Corresponding author adrian.irles@ific.uv.es}}

\affil{IFIC, Universitat de Val\`encia and CSIC, C./ Catedr\'atico Jos\'e Beltr\'an 2, E-46980 Paterna, Spain}

\begin{document}

\maketitle

\begin{abstract}
Highly compact and finely segmented silicon–tungsten electromagnetic calorimeters are being developed within the FCAL collaboration for applications in the LUXE experiment at DESY and future electron–positron collider facilities. 
These detectors combine tungsten absorber plates with thin silicon pad sensors, providing a small effective Moli\`ere radius and high spatial granularity, which are essential for resolving nearby electromagnetic showers in high-occupancy environments.

The fundamental active unit of this calorimeter concept is the Compact Silicon Sandwich (CSIS), 
integrating a silicon pad sensor together with signal routing, high-voltage distribution and mechanical support in a highly compact structure. The assembly of these CSIS modules is performed within a dedicated infrastructure for silicon detector integration.

A partially instrumented prototype of such a calorimeter has been tested in an electron beam with energies between 1 and 6~GeV. First results from the 2025 test beam campaign are presented, including minimum-ionizing particle calibration and preliminary event displays illustrating the shower development in the highly granular detector. These results constitute an important step towards the validation of this technology for LUXE and future collider experiments.
\end{abstract}

\vfill
\begin{center}
\textit{Contribution to the International Workshop on Future Linear Colliders (LCWS 2025), 20-24 October 2025. Valencia, Spain (C25-10-20.1)}
\end{center} 
\clearpage



\section{Introduction}

Highly compact and finely segmented electromagnetic calorimeters are essential for a variety of applications in future particle physics experiments. In particular, they are required for precision luminosity measurements at electron--positron colliders~\cite{Behnke:2013xla,CEPCStudyGroup:2023quu,FCC:2025lpp,LinearCollider:2025lya} and for the measurement of electron and positron multiplicities and energy spectra in the LUXE~\cite{LUXE:2023crk} experiment at DESY and the European XFEL, which investigates strong-field quantum electrodynamics.

These applications require detectors combining a small effective Moli\`ere radius with high transverse granularity, enabling precise reconstruction of electromagnetic showers.

In this contribution, the design and prototyping of a highly compact silicon--tungsten electromagnetic calorimeter are presented. Particular emphasis is placed on the Compact Silicon Sandwich (CSIS\footnote{Pronounced ``ce-sis'', as commonly used by Spanish speakers.}) modules and their integration within the readout and mechanical structure, as well as on the first results from the 2025 test beam campaign.


\section{Compact silicon--tungsten calorimetry for LUXE and Higgs Factories }

The concept of compact silicon--tungsten sampling calorimeters has been developed within the FCAL collaboration for forward calorimetry at future collider experiments. In the context of the LUXE experiment, this detector is referred to as ECALp, as it is designed for the measurement of positrons produced in the laser--electron interaction.

These developments are closely connected to the DRD Calorimetry (DRD Calo) programme, which focuses on generic R\&D for calorimeter technologies, while FCAL targets their implementation in concrete detector systems such as LUXE and luminometers for future Higgs factories.

The calorimeter consists of tungsten absorber plates interleaved with thin active layers implemented CSIS modules.

The overall detector performance relies on both the CSIS integration and dedicated front-end electronics. The present prototype is based on the FLAME ASIC~\cite{Firlej:2026ntf}, while the full readout architecture is described in detail in a companion contribution to these proceedings~\cite{ECALp_readout}.

The mechanical structure ensures precise positioning of the absorber plates and active layers under stringent compactness constraints. Achieving this requires sub-millimetre control of layer thickness and planarity. The stack is assembled using precisely machined tungsten plates and a lightweight support structure to guarantee alignment and reproducibility of the detector response. 


\section{Compact Silicon Sandwich, CSIS}

The fundamental active unit of the calorimeter is the CSIS, which integrates the sensitive detector element together with the mechanical and electrical interfaces required for operation within a highly compact tungsten--silicon sampling structure.

\begin{figure}[htbp]
\centering
\includegraphics[width=0.38\textwidth]{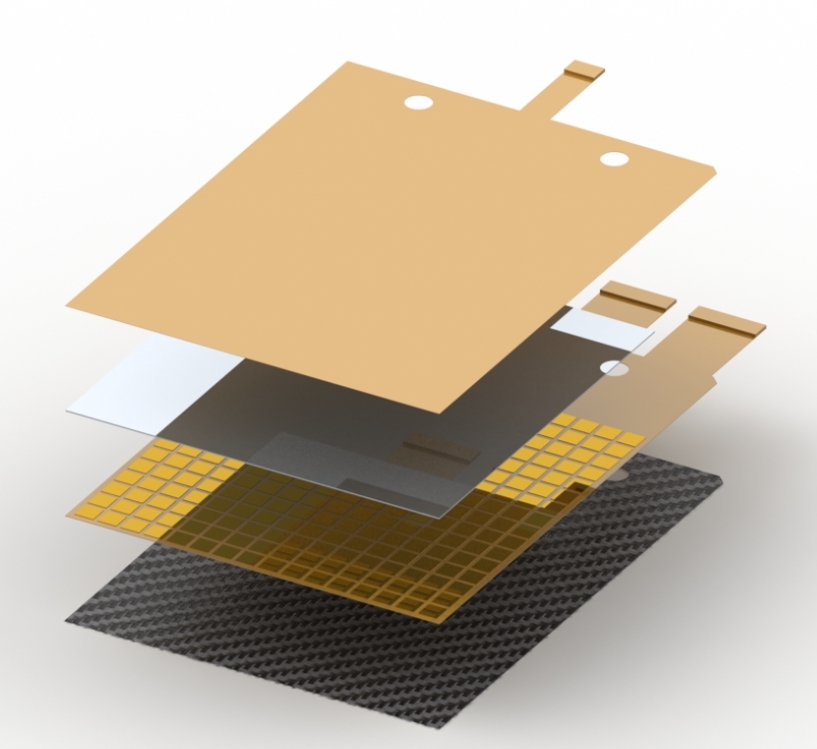}
\includegraphics[width=0.48\textwidth]{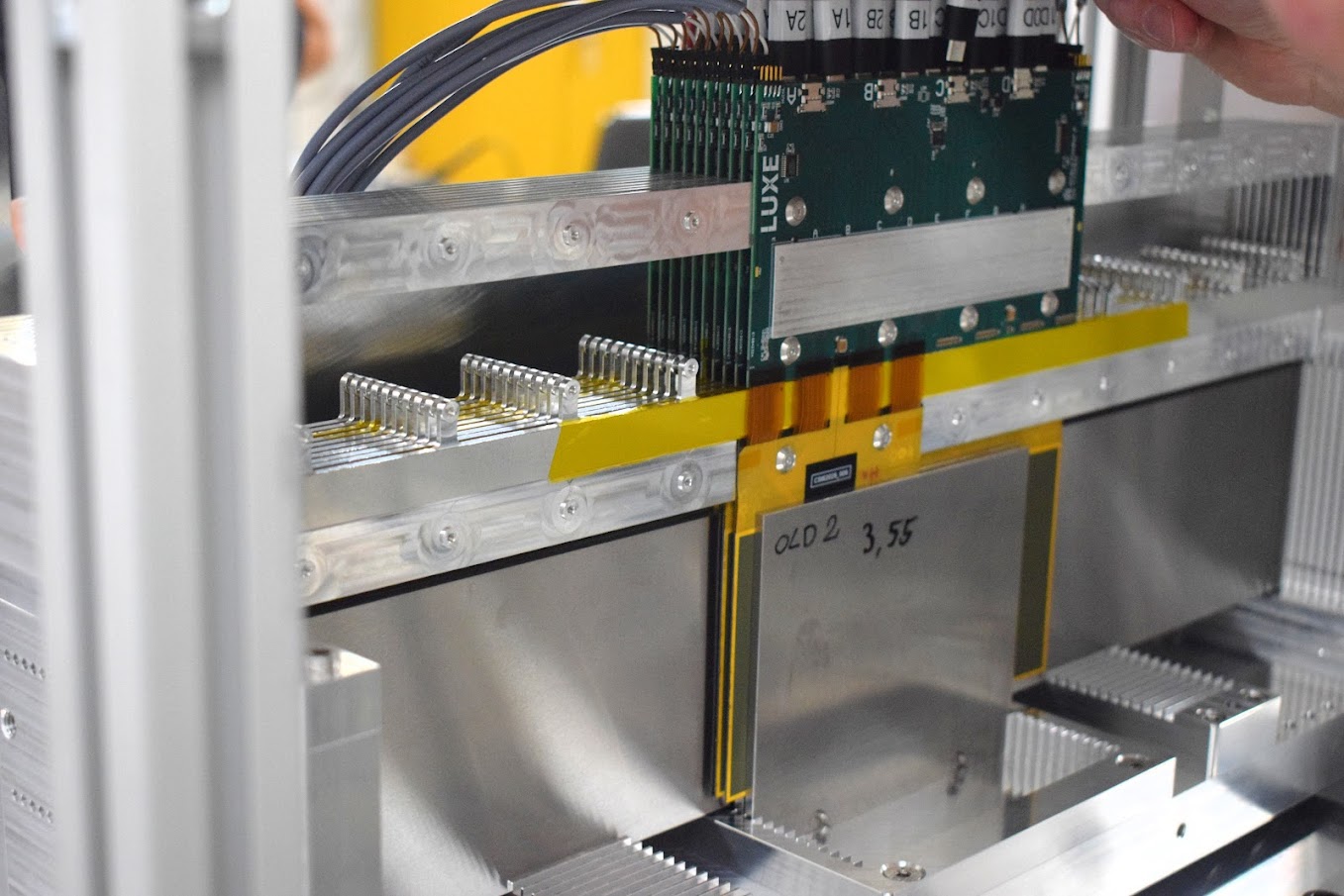}
\caption{Left: schematic layout of a CSIS module, showing (from top to bottom) the HV foil, silicon pad sensor, fan-out routing and the carbon-fiber sheet for mechanical support. Right: photograph of the mechanical structure with nine towers of two CSIS inserted between long tungsten plates and a tenth tower made of only one CSIS (partially visible) inserted in the front. The setup was completed later with one more CSIS in front of the first one plus additional plates of tungsten.}
\label{fig:CSIS}
\end{figure}

Each CSIS module is based on a silicon pad sensor of approximately $90\times90$~mm$^{2}$, segmented into $16\times16$ pads with a pitch of about $5.5\times5.5$~mm$^{2}$.  
The sensor is complemented by a flexible Kapton fan-out for signal routing and a high-voltage (HV) foil providing a uniform bias. All components are embedded in a thin carbon-fibre support structure ensuring mechanical stability while preserving compactness. An expanded view of the design of the CSIS can be seen in the left part of Figure \ref{fig:CSIS}.

A key requirement of the CSIS concept is the strict control of the total thickness of the active layer, which must remain below $\sim$\,1~mm to minimise the effective Moli\`ere radius. In the current prototype used in the 2025 test beam campaign, this constraint was slightly relaxed to about 1.2~mm, while maintaining the overall compactness of the calorimeter. A partially instrumented ECALp can be seen in the right part of Figure \ref{fig:CSIS}, where the first CSIS is partially visible. The setup is described in Section \ref{sec:TB}.

\subsection{Sensor integration and signal routing}

The electrical connection between sensor pads and fan-out traces is achieved using conductive adhesive, avoiding wire bonding in the active area and reducing the material budget. The routing is optimised to ensure uniform response and minimise parasitic effects.

The front-end electronics are located outside the sensitive region, allowing the active area to remain free of additional material. A detailed description of the front-end electronics and data acquisition system is presented in a separate contribution to these proceedings.

The integration of multiple CSIS units forms the active layers of the calorimeter, enabling a scalable and highly granular detector architecture.

\subsection{Module assembly and characterisation}

The assembly of the CSIS modules involves the precise integration of the silicon sensor, fan-out, HV foil and mechanical support. Particular care is required to control the planarity and alignment of the components, as well as to ensure reliable electrical connections and stable operation under bias voltage.

The production and assembly of the modules are carried out at the TARDIS Lab at IFIC, a dedicated infrastructure for silicon sensor characterisation and detector module integration~\cite{TARDIS_proceeding}. The CSIS modules undergo electrical and mechanical characterisation to validate their performance and compliance with the design requirements. A detailed description of the assembly procedures and characterisation results is given in a companion contribution to this proceedings.


\section{Test beam campaign 2025}
\label{sec:TB}
A prototype of the highly compact silicon--tungsten calorimeter was tested in an electron beam at DESY in 2025. Electron beams with energies between 1 and 6~GeV were used, and several million events were recorded during the campaign. A telescope made of 6 planes of ALPIDE monolithic active pixel sensors was employed~\cite{ALPIDETelescope} to measure the trajectory of each incoming electron, providing precise tracking information for the analysis.

The calorimeter prototype consisted of a stack of tungsten absorber plates interleaved with CSIS modules, forming a partially instrumented sampling calorimeter. The signal on each pad was amplified and digitised by front-end electronics based on the FLAME ASIC~\cite{Firlej:2026ntf}. The setup included nine layers instrumented with two sensors arranged in adjacent towers, complemented by two additional layers equipped with a single sensor in the front region, where the transverse shower profile is narrower. In total, the prototype comprised 20 CSIS units, corresponding to 5120 pads. 
As part of the prototype configuration, the absorber structure includes plates of different transverse dimensions and slightly reduced density, corresponding to tungsten alloys. These variations are taken into account in the detector description used for simulation and analysis. Due to the limited availability of front-end ASICs, only a subset of these channels was instrumented with the FLAME readout, resulting in approximately half of the pads being actively readout during the test beam campaign.

\begin{figure}[h!tbp]
 \begin{center}
  \includegraphics[width=0.5\textwidth]{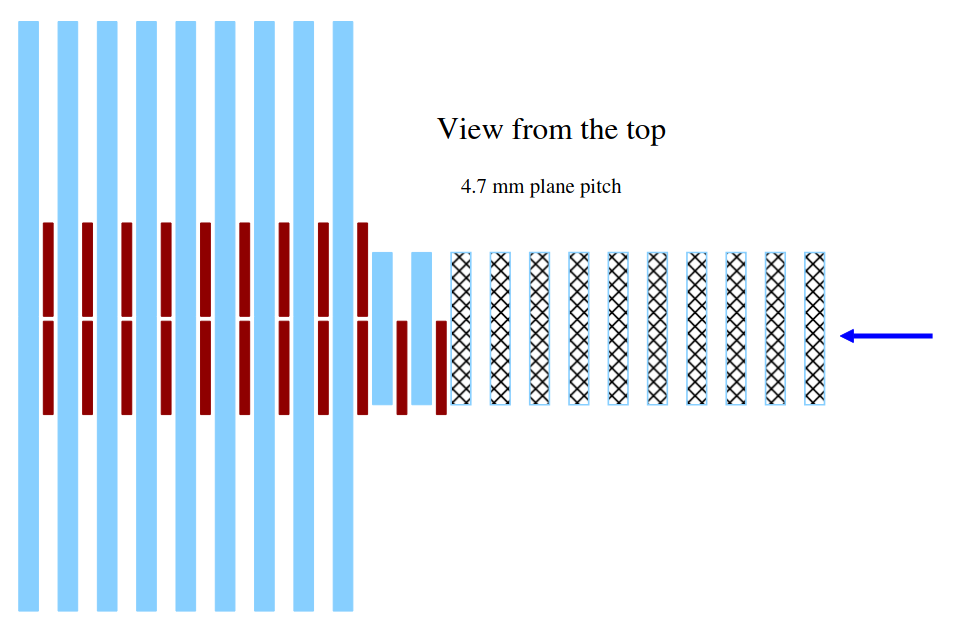}  \\\includegraphics[width=0.5\textwidth]{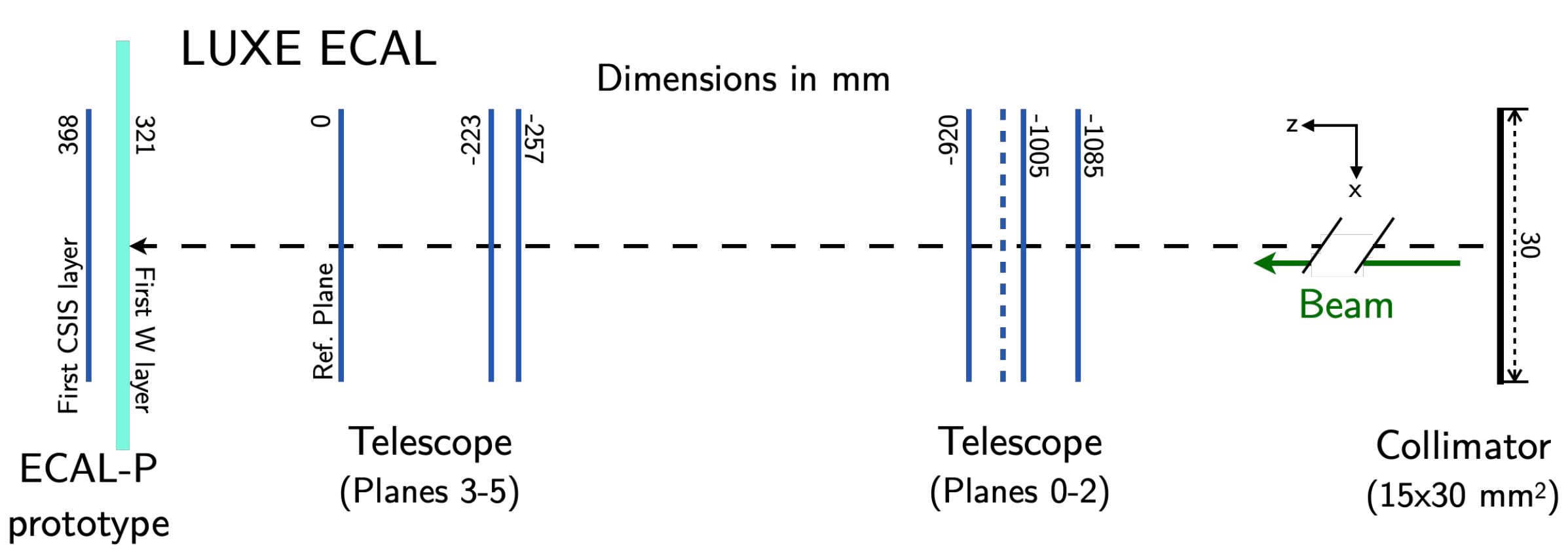}
 
    \caption{
    Sketch of the test beam set-up in 2025. Upper figure: expanded view of the calorimeter (not at scale). Blue and dashed rectangles represent the different tungsten layers, with the dashed ones being the ones corresponding to the extension of effective depth of the calorimeter, as described in the text. The 20 CSIS are represented by the dark-red block. Lower figure: view of the telescope setup and the first layer of the ECALp. Electrons arrive from the right, pass the first scintillator, the first group of three telescope sensors, then the second scintillator and the second group of another three telescope sensor planes, and hit the ECALp prototype. The dimensions are given in mm.} 
    \label{fig:TB_set-up}
    \end{center}
\end{figure}

To extend the effective depth of the calorimeter and study electromagnetic shower development up to about $20$ radiation lengths ($X_0$), additional tungsten absorber material was placed in front of the instrumented region. As part of the prototype configuration, the absorber structure includes plates of different transverse dimensions and slightly reduced density, corresponding to tungsten alloys. These variations are taken into account in the detector description used for simulation and analysis. A schematic view of the setup is shown in Figure~\ref{fig:TB_set-up}.

\subsection{First results}

The raw data are corrected for pedestal variations and common-mode noise prior to channel calibration. The treatment of common noise is essential to achieve a stable detector response. A detailed description of the data processing chain and readout system is provided in Ref.~\cite{ECALp_readout}.

\begin{figure}[htbp]
\centering
\includegraphics[width=0.48\textwidth]{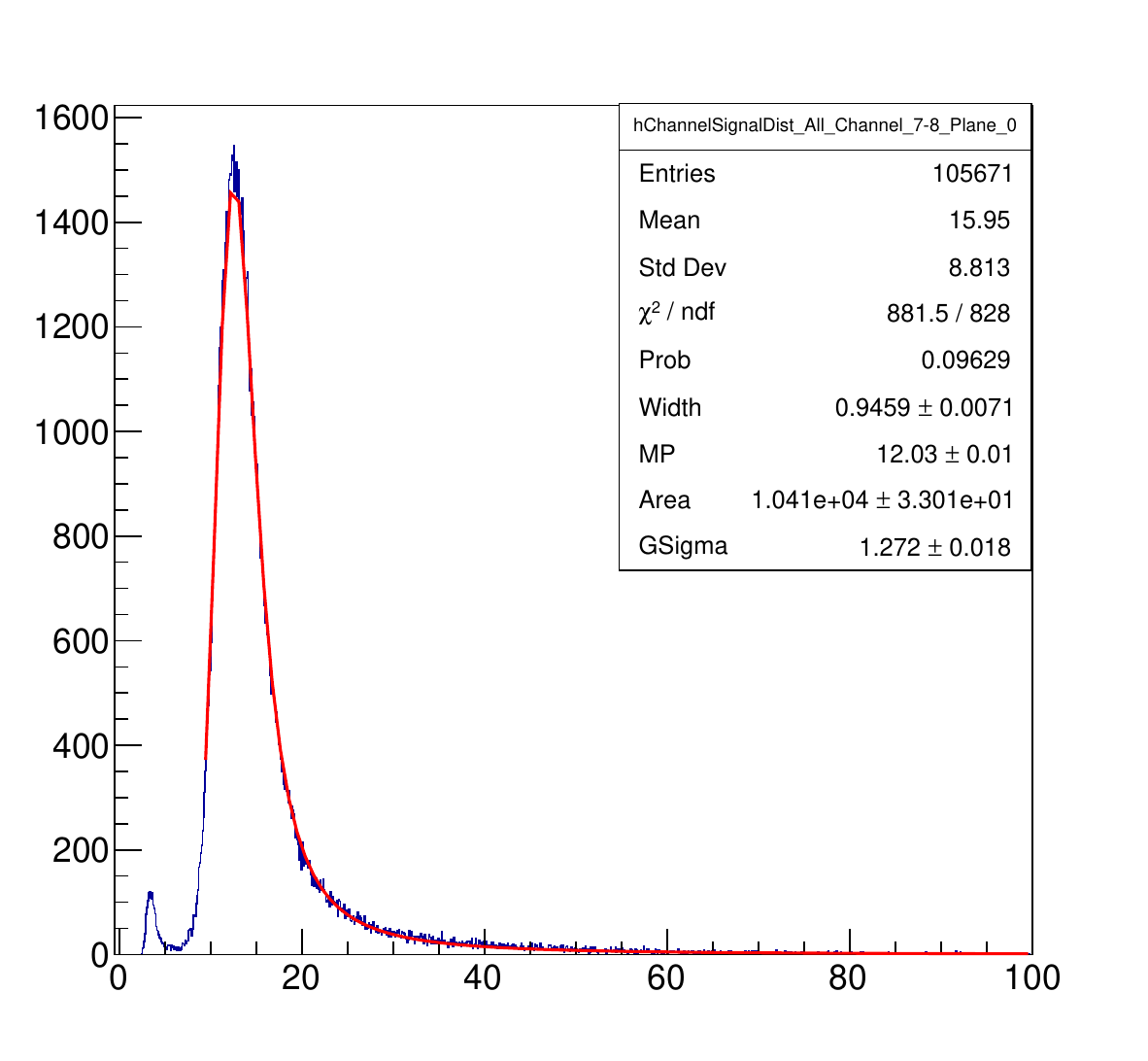}
\includegraphics[width=0.48\textwidth]{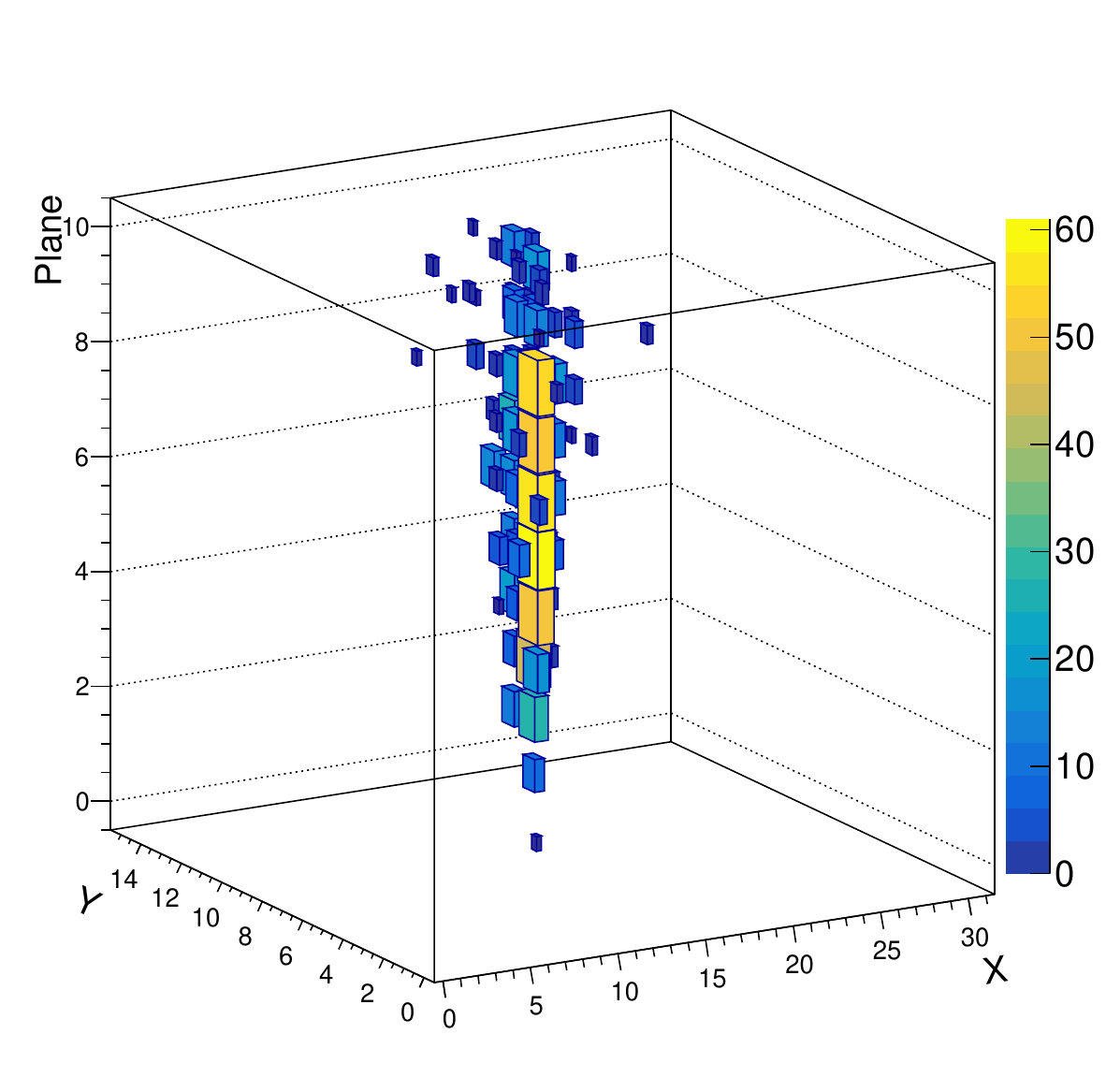}
\caption{Left: example of the signal distribution measured in a single pad for minimum-ionizing particles. The distribution is fitted with a Landau function to extract the most probable value used for calibration. 
Right: preliminary event display of an electromagnetic shower recorded in the calorimeter prototype during the 2025 test beam campaign. The colour scale represents the signal amplitude in each pad.}
\label{fig:MIP_event_display}
\end{figure}

The response of the detector to minimum-ionizing particles (MIPs) provides a reference for the calibration of individual channels. Figure~\ref{fig:MIP_event_display} (left) shows an example of the signal distribution recorded in a single pad, exhibiting the characteristic Landau distribution. 

After data processing, a signal-to-noise ratio for MIPs at the level around 10 is observed, demonstrating a clear separation between signal and noise. The distribution of the calibrated MIP response across channels shows a good level of uniformity, with variations near 5\% over all channels.

The high granularity of the detector allows a detailed imaging of electromagnetic shower development. Figure~\ref{fig:MIP_event_display} (right) shows a preliminary event display of an electron-induced shower in the calorimeter prototype.

These first observations illustrate the capability of the detector to resolve the spatial structure of electromagnetic showers. A detailed quantitative analysis of the detector performance is ongoing and will be reported in a dedicated publication.


\section{Outlook}

The data collected during the 2025 test beam campaign are currently under detailed analysis. Ongoing work includes detector commissioning studies, calibration refinement and alignment procedures based on the pixel telescope, as well as improvements in track reconstruction and its integration with the calorimeter response.

The goal of this programme is to extract the key performance parameters of the detector, such as energy resolution, linearity and effective Moli\`ere radius. In addition, dedicated studies exploiting the high granularity of the detector are being pursued, including the detailed characterisation of shower development and the impact of inactive regions on the detector response.

These studies will provide a comprehensive validation of the highly compact silicon--tungsten calorimeter concept for its application in the LUXE experiment and future Higgs factory detectors. The results will be presented in forthcoming publications.


\section*{Acknowledgements}
\addcontentsline{toc}{section}{Acknowledgements}

This research was supported by the European Union’s Horizon 2020 Research
and Innovation programme 3309 under GA no. 101004761; the Weissfeld Family Charitable Foundation, the Israeli PAZY Foundation (ID 318);
the Spanish MICIU/AEI and European Union/FEDER via the grant \texttt{PID2024-158190NB-C21} the Generalitat Valenciana (GV) via  PlanGenT program with the grant number \texttt{CIDEGENT/2020/021};
the National Science Centre, Poland, under grant no. \texttt{2021/43/B/ST2/01107};
University of Warsaw Excellence Initiative IDUB grant no. \texttt{SP-501-D113-20-0001380};
the Romanian Ministry of Research, Innovation and Digitalization under Romanian National Core Program LAPLAS VII contract no. \texttt{30N/2023} and by the Institute of Atomic Physics through the CERN-RO Project \texttt{No.005/2024}.

The measurements leading to these results have been performed at the Test Beam Facility at DESY Hamburg (Germany), a member of the Helmholtz Association (HGF).

\printbibliography

\end{document}